\documentclass[12pt,a4paper]{article}
\usepackage{epsf}
\usepackage[francais]{babel}
\usepackage[utf8]{inputenc}  
\usepackage[T1]{fontenc}
 \usepackage{graphics}
\usepackage{natbib}
\usepackage{placeins}
\newcommand{\be}{\begin{equation}}
\newcommand{\nd}{\end{equation}}

\begin{document}

\title{Persistence of discrimination:
revisiting Axtell, Epstein and Young.}
\author{
        Gérard Weisbuch \\
 Ecole normale superieure, 24, rue Lhomond, Paris, France   \\
Laboratoire de physique statistique, Département de physique de l’ENS, \\École normale supérieure, PSL Research University, Université Paris Diderot, \\
Sorbonne Paris Cité, Sorbonne Universités, UPMC Univ. Paris \\
06, CNRS, 75005 Paris, France}

\maketitle

email: weisbuch@lps.ens.fr


\date

\begin{center}
  Abstract

We reformulate an earlier model of the "Emergence of classes..."  
proposed by \cite{aey} using more elaborate cognitive
processes allowing a statistical physics approach.
The thorough analysis of the phase space and of the basins of attraction 
leads to a reconsideration of the previous social interpretations: our model 
predicts the reinforcement of discrimination biases and their long term 
stability rather than the emergence of classes. 
\end{center}

 \section{Introduction}

 During the 90's social scientists introduced
several thought provocative models of social phenomena, most often 
using numerical simulations (multi-agent simulations).
These models have later been extended by methods and concepts derived from
statistical physics such as Master Equations and Mean Field Approximation. A few examples include voters models and imitation processes
\cite{andre} and the review of \cite{rmp} ,
El Farol and the minority game \citep{El-Farol} and \cite{challet}, diffusion of cultures \citep{axel} and \citep{castel}.
 Revisiting these models provided deeper insight, more precise results
 and even sometimes corrections. 
 
   The questions of the emergence and persistence of classes and discrimination
   received a lot of attention from social scientists, ethnographers and 
   economists, see e.g. \citep{bowles} and references within.
   A very inspiring model entitled "Emergence of Classes in a
Multi-Agent Bargaining Model" was
   proposed by  Axtell, Epstein and Young (\cite{aey}). We here propose to
   revisit their approach using a more elaborate model of agent cognition
   and to compare a mean field approach to our agent based simulation results.

\section{The models}
  \subsection{The original model of Axtell, Epstein and Young}
  Let us briefly recall the original hypotheses and the main results of 
Axtell, Epstein and Young (\cite{aey}).
  \begin{itemize}
    \item Framework: pairs of agents play a bargaining game introduced
    by \cite{Nash} and \cite{PY}. During sessions of the game, each agent 
    can, independently of his opponent, request one among three demands:
    L(ow) demand 30 perc. of a pie, M(edium) 50 perc. and H(igh) 70 perc. 
    As a result, the two
    agents get at the end of the session 
    what they demanded when the sum of both demands is less
    than the 100 perc. total; otherwise they don't get anything.
    The corresponding payoff matrix is written table (1).
    At each step a random pair of agents is selected to play the bargaining
    game. The iterated game is played for a large number of sessions, much larger that 
    the total number of agents which could then learn from their experience
      how to improve their performance.
    \item Learning and memory:  Agents keep records of the previous demands of 
    their opponents, e.g. for 10 previous moves.
    \item Choosing the next move: at each time step, pairs of agent are randomly 
    selected to play the bargaining game.  They most often
     choose the move that optimises their expected payoff
 using the memory of previous encounters as a surrogate for the actual probability distribution of their opponent's next moves.
  With a small probability $\epsilon$,
      e.g. 0.1, they choose randomly among L, M, H. 
    \end{itemize}  
    
  The main results obtained by \cite{aey} from numerical simulations are:
  \begin{itemize}
  \item  They observe different transient configurations which they 
  interpret as "norms", e.g. the equity norm is observed when all agents
  play M. Because of the constant probability of random noise, the system never stabilises on an attractor, even in the sense of Statistical Physics.
  The duration of the transients increases exponentially with
  the memory size and $1/\epsilon$.
  \item Their most fascinating result is obtained when agents are divided
  into two populations with arbitrary tags say e.g. one red and one blue.
  When agents take into account tags for playing and memorising games
  (in other words when agents play separately two games, one intra-game
   against agents with the same tag and another inter-game against agents
   with a different tag)
  one observes configurations in the inter-game such that one population always play H while the other population plays L; they interpret such inequity norm as the emergence of classes, the H playing population being 
  the upper class.  
  \end{itemize}
  
  Equivalent results are obtained when agents are connected via a social network as observed by \citep{poza}
  on a square lattice as opposed to the full connection structure used by \citep{aey}. For some instances, domains with different norms
   occupy different parts of the lattice. 
  Otherwise, one single domain of agents playing the same norm
  covers the entire lattice, depending upon the initial conditions. 
   
  \vspace{1cm}
  From now on, we follow a plan starting with the exposition of
  our own model (section 2.2). 
   The use of a mean field approximation
   allows to simply describe the attractors of the dynamics and 
  the different dynamical regimes (section 3). These results are then compared with 
  those obtained by direct agent based simulations (section 4), including a thorough survey of the attraction basins. We further proceed with the analysis
  of the two tagged populations version (section 5). 
  The discussion compares our results
  to those of previous models and to magnetic systems.
  A short conclusion stresses the difference
  in interpretation of the models in terms of social phenomena (section 6). 
  
  \subsection{The moving average and Boltzman choice cognitive model}
  We start from the same bargaining game as \citep{aey}
   with a payoff matrix written in table (1),
  but using different coding of past experience 
  (moving average of past profits) and choice function (Boltzman function).
  
\begin{table}
\centerline{\begin{tabular}{|r|r|r|r|}
\hline
& L&M&H\\
\hline
L&0.3&0.3&0.3\\
\hline
M&0.5&0.5&0\\
\hline
H&0.7&0&0\\
\hline
\end{tabular}}
\caption {Payoff matrix of the Nash demand game. The first column 
represents the move of the first player L, M or $H$. The first row represents
 the move of her opponent. The figures in the matrix represent 
the payoff obtained by the first player.}
\end{table}

  
 The present model is derived from standard models of reinforcement learning in cognitive science, see for instance \cite{mars}. 
  
  Rather than memorising a full sequence 
  of previous games, agents update 3 "preference coefficients" $J_j$
for each possible move $j$, based on a moving average of the profits they
made in the past when playing $j$. $J_1$ is the preference coefficient
for playing $H$, $J_2$ 
for $M$ and  $J_3$ for $L$. 
 The updating process following time interval $\tau$ 
after a transaction is: 
  \be
 J_{j}(t+\tau)  = (1-\gamma) \cdot J_{j}(t) + \; \pi_{j}(t) , 
 \qquad \forall j, \label{J1}
 \nd
 The decrease term in $1-\gamma$ corresponds to discounting the importance of past
 transactions, which makes sense in an environment varying with 
 the choices of the other players. $\pi_{j}(t)$ is the actual profit 
 made during the chosen transaction $j$; the 2 other $J_{j'}$ corresponding to 
 the 2 other choices $j'$ are simply decreased.
 
   These preference coefficients are then used to choose 
   the next move in the bargaining game. Agents face an
    exploitation/exploration dilemma: they can decide to exploit the information
   they earlier gathered by choosing the move with the highest 
   preference coefficient or check possible evolutions of profits    
by trying randomly other moves. Rather than using a constant rate
of random exploration $\epsilon$ as in \cite{aey}, the probability of choosing demand
$j$ is based on the logit function:
 \be
 P_{j}  =   \frac{\exp (\beta J_{j})}{\sum_{j}{\exp (\beta J_{j})}} , 
 \qquad  \forall j, \label{P}
 \nd
  where $\beta$, the discrimination rate, 
 measures the non-linearity of the relationship between the
 probability $ P_{j}$  and the preference coefficient $ J_{j}$. 
 Large $\beta$ values results in always playing the choice $j$ 
 with the largest $J_j$,
 small  $\beta$ values to indifference among the three choices.
 Economists use the name logit for the Boltzmann distribution. 
 We have earlier shown \cite{nadal} that the Boltzmann distribution 
 can be derived by maximising a linear combination of expected profits and 
 information gained through exploration, see \citep{bouchaud} for a thorough discussion. 
 

  Comparing our model with the one proposed by \citep{aey}:
\begin{itemize}
  \item The moving average corresponds to a gradual rather than abrupt
  decrease of previous memories, it is based on agent's own experience in terms of profit rather than the observation of her opponents' moves and it uses less memory.
  \item Boltzman choice has a random character as the constant probability noise introduced in \citep{aey}, but furthermore the choice depends upon 
  the differences in experienced profits; we might expect agents to be less hesitant
  when their previous experience resulted in very different preference coefficients. 
  \end{itemize}

 \section{The mean field approximation}
 
 \subsection{Derivation of the mean field approximation}
 
 The difference equation (1) can be changed to a differential 
 equation in the limit of a slow dynamics: 

\be
\frac{dJ_{j}}{dt} \; = \; - \gamma J_{j}(t) + \pi_{j}(t)
\nd
 where the time unit is the average time between the agent's bargaining processes.
 $\pi_{j}(t)$ is the profit made by the agent if he chose demand $j$.

The Mean Field approximation
consists in replacing $\pi_{j}(t)$
by its expected value $<\pi_{j}>$, thereby transforming the stochastic
differential equation into a deterministic differential equation.

The time evolution of $J_{ij}$ is thus approximated by the following set 
of equations:
\be
\frac{dJ_{j}}{dt} = - \gamma J_{j} + <\pi_{j}>
\label{diff}
\nd
where $<\pi_{j}>$ is given by:
\be
<\pi_{j}>  =  \frac{\sum_{i} \pi_{ij}\exp(\beta J_{i})}{\sum_{i}
\exp(\beta J_{i})};
\nd
 $j$ is the agent's move, $i$ are the 3 possible moves of her opponent,
 and the $\pi_{ij}$ (0, 0.3, 0.5, 0.7) are the coefficients of the pay-off matrix.
 The mean field approximation neglects fluctuations among agents representations,
 their $J_j$. Hence agent $j$ evaluates the probability of her opponent's moves
 according to her own estimations, using Boltzman functions of her own $J$.


Using statistical physics notation $Z$:
\be
Z = \sum_{i} \exp(\beta J_{i})
\nd
the internal representation of the agent is thus vector 
($J_1,J_2,J_3$) which components obey dynamics:
\be
 \frac{dJ_{1}}{dt} = - \gamma J_{1} + (exp(\beta J_{1})*0.7*exp(\beta J_{3}))/Z/Z
\nd
\be
 \frac{dJ_{2}}{dt} = - \gamma J_{2} + (exp(\beta J_{2})*0.5(exp(\beta J_{2})+exp(\beta J_{3}))/Z/Z
\nd
\be
\frac{dJ_{3}}{dt} = - \gamma J_{3} + exp(\beta J_{3})*0.3/Z
\nd

 Taking the exponentials as new variables simplifies expressions (7-9)
 and allows to deduce scaling properties. 
Let:
\be
x=exp(\beta J_{1}),  \; J_{1}=log(x)/\beta
\nd
\be
y=exp(\beta J_{2}),   \;  \;  \; J_{2}=log(y)/\beta
\nd
\be
z=exp(\beta J_{3}),  \; \; J_{3}=log(z)/\beta
\nd
The new equations are:
\be
 \frac{dx}{\beta dt} = x  (- \alpha  log(x) + 0.7  \frac{x z}{s^{2}})
\nd
\be
 \frac{dy}{\beta dt} =  y (- \alpha  log(y) + 0.5  \frac{y (y+z)}{s^{2}})
\nd
\be
\frac{dz}{\beta dt} = z (- \alpha  log(z) + 0.3  \frac{z}{s})
\nd
with $s=x+y+z$. 

Expressions  (13-15) show that a single parameter $\alpha=\frac{\gamma}{\beta}$
 determines equilibrium conditions, an improvement on \citep{aey} who
 needed two parameters $\epsilon$ and memory size. 
 Phase transition diagrams will then be drawn varying $\beta$ while keeping 
 $\gamma=0.05$ constant.
$\beta$ plays the role of a kinetic coefficient,
increasing the characteristic time towards equilibrium.
The magnitude of the $J$ coefficients at equilibrium scales
as $1/\gamma$.
  
  


  \subsection{Mean field analysis: Attractors and transitions}   
  

  The state of the system is described by the set of the preference
  coefficients $J_1$, $J_2$ and $J_3$ of the agents, 
  i.e. their estimated profit divided by $\gamma$ for the three possible moves resp. H, M and L. This is an improvement with respect to \citep{aey}
  which space phase dimension was three times the memory size. 
 Our analysis can then proceed using the more powerful methods of dynamical systems and statistical mechanics rather the Markovian formulation proposed in 
  \citep{aey}. 
  
 Trajectories in the {\bf J} phase space are obtained
  by solving the mean field equations (4-5) using a Rosenbrock integrator \citep{GRIND}.   
  Grids of trajectories help to figure out attractors and attraction basins.  
   Since we cannot draw sets of 3 trajectories, we display their projections 
  in plans ($J_1,J_2$) , ($J_1, J_3$) and ($J_3,J_2$) for a given choice of
   $\beta=2,\gamma=0.05$ in figure (1).
    The trajectories start at regular interval in the projection plan 
     with the same third $J$ coordinate.   



\begin{figure}
\epsfxsize=80mm\epsfbox{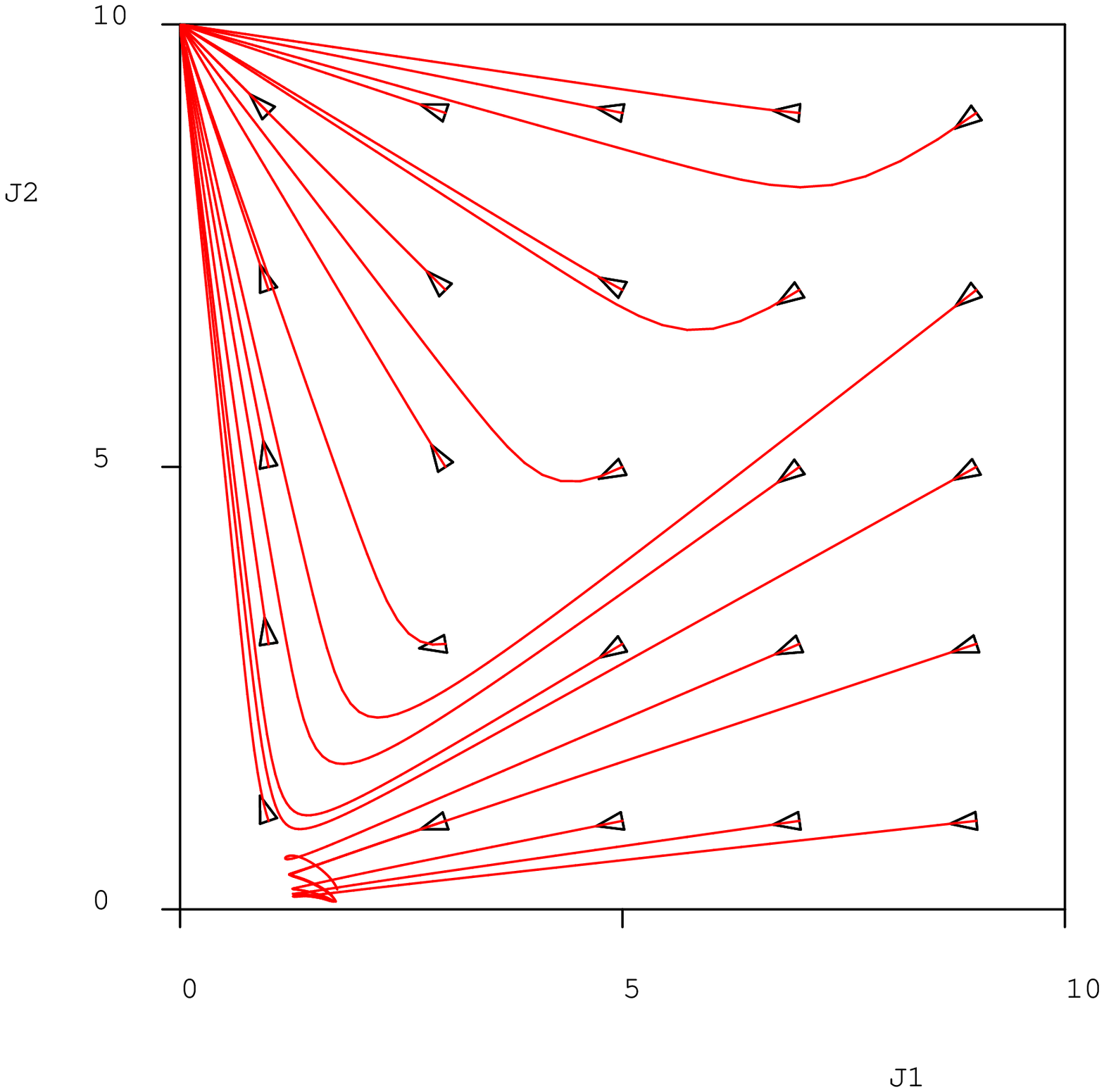} \epsfxsize=80mm\epsfbox{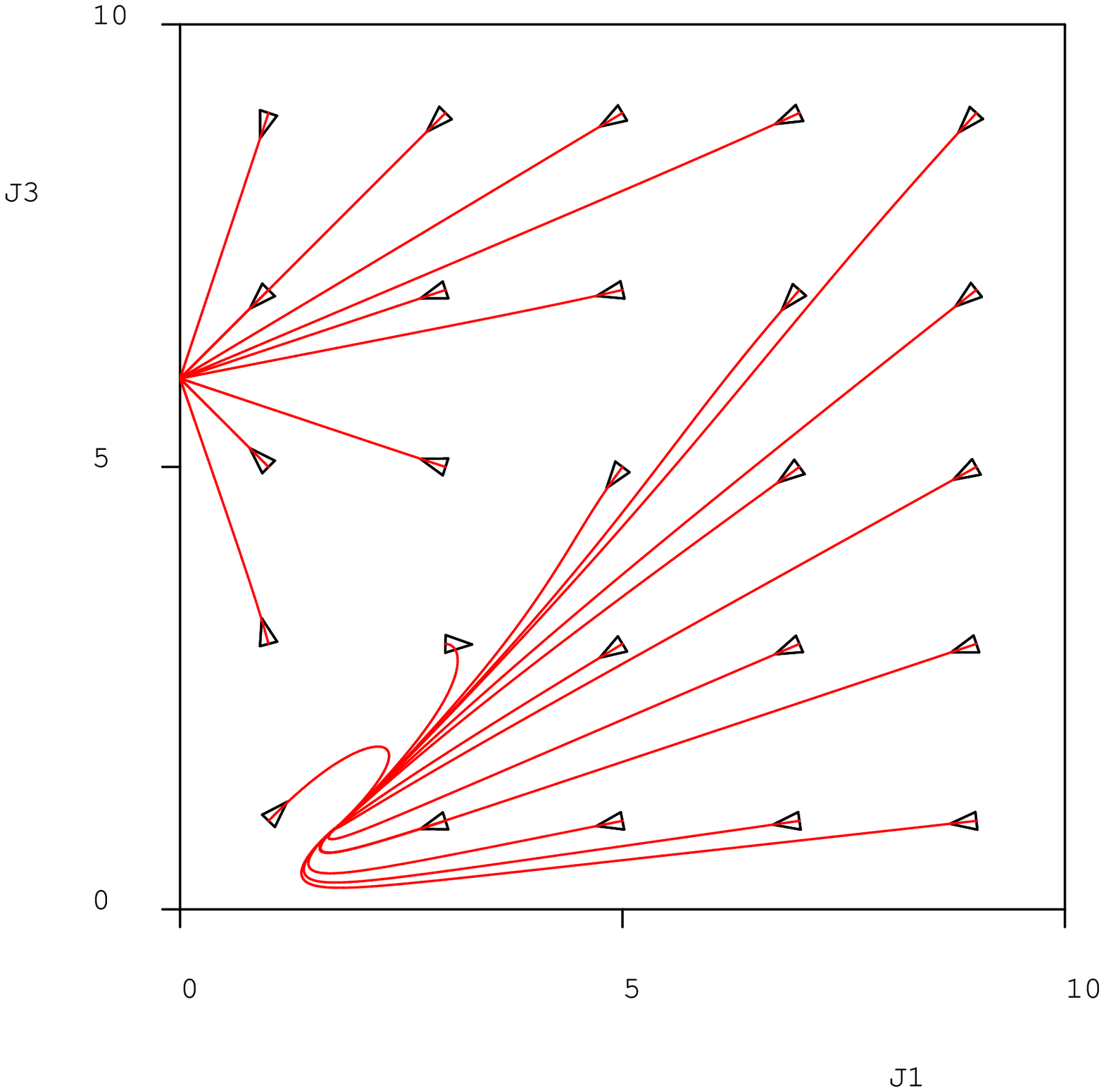} \epsfxsize=80mm\epsfbox{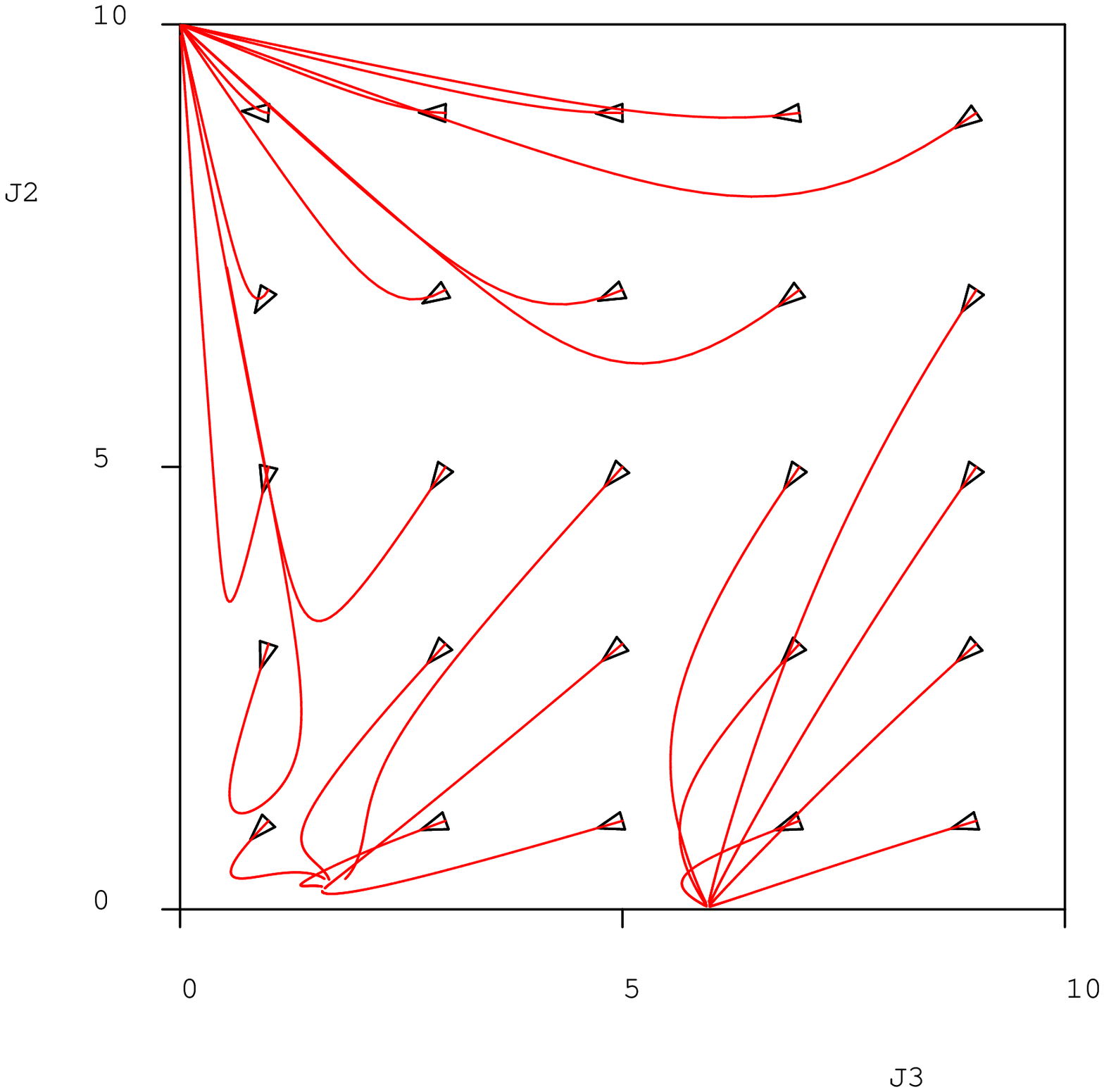} \epsfxsize=80mm\epsfbox{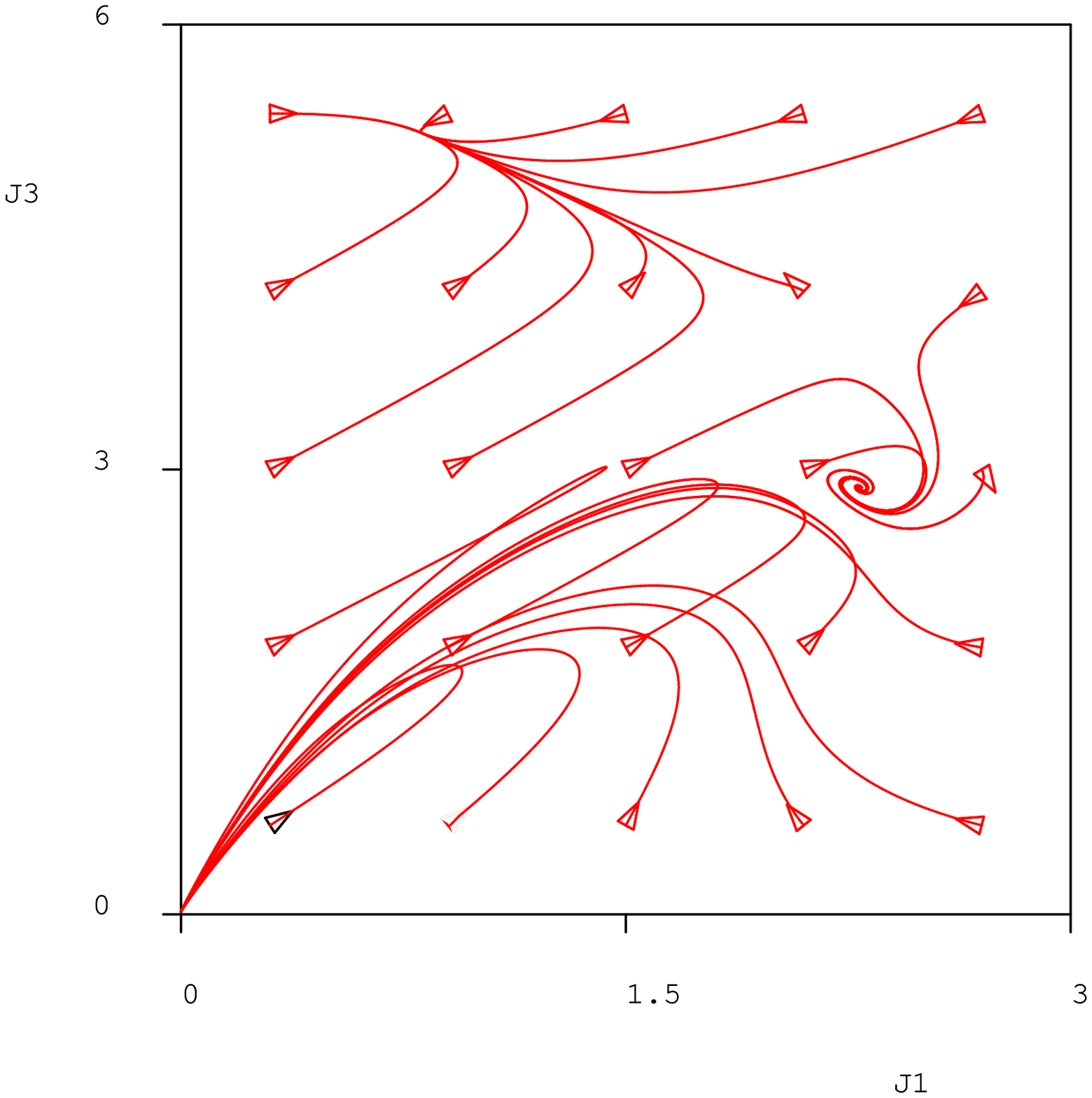} 
\caption{ Grids of trajectories projected on the ($J_1,J_2$) plan
starting from $J_3=5$ (above left),
on the ($J_1, J_3$) plan starting from $J_2=0.5$ (above right)
and on the ($J_3, J2$) plan starting from $J_1=6$ (below left).
The black triangles figure the initial conditions of the
individual trajectories.
 $\beta=2, \gamma=0.05$ . 
 The fourth set of projections on the ($J_1,J_3$) plan
  starting from $J_2=0.5$ (below right) was obtained
just above the bifurcation for $\beta=0.58.$ and $\gamma=0.05$ .
The L and HL attractors get closer and the basin of attraction of the HL attractor is strongly reduced by the widening of the M attractor. \cite{GRIND} software.
}
\end{figure}

  Three attractors can be observed:
  one with large $J_2$ when move M is the preferred choice by all agents,
  to be called the M attractor;
   one with large $J_3$ when move L is the preferred choice by all agents,
   to be called the L attractor;
   and one with lower values of $J_1, J_2$ and $J_3$, to be called the HL attractor. 
   
   The Mean Field analysis readily tells us that two of the attractors
   are such that all agents always play the same strategy
   either M or L.  
   
     The dependence of the J's upon the reduced parameter $\beta$/$\gamma$ 
     is displayed on the continuation plot of figure 2.
     We clearly identify the 3 same attractors in the ordered regime above 
     $\beta=0.57$, and only one attractor left with move M as the preferred choice
     for lower $\beta$ values $0.3 < \beta < 0.57$.
     The bifurcation is observed around $\beta \approx 0.57$.
     A steep, but not abrupt, transition further
    occurs when $\beta \approx 0.3$
     to a disordered regime such that agents do not display strong
     preferences for any choice.
     
\begin{figure}
\centerline{\epsfxsize=100mm\epsfbox{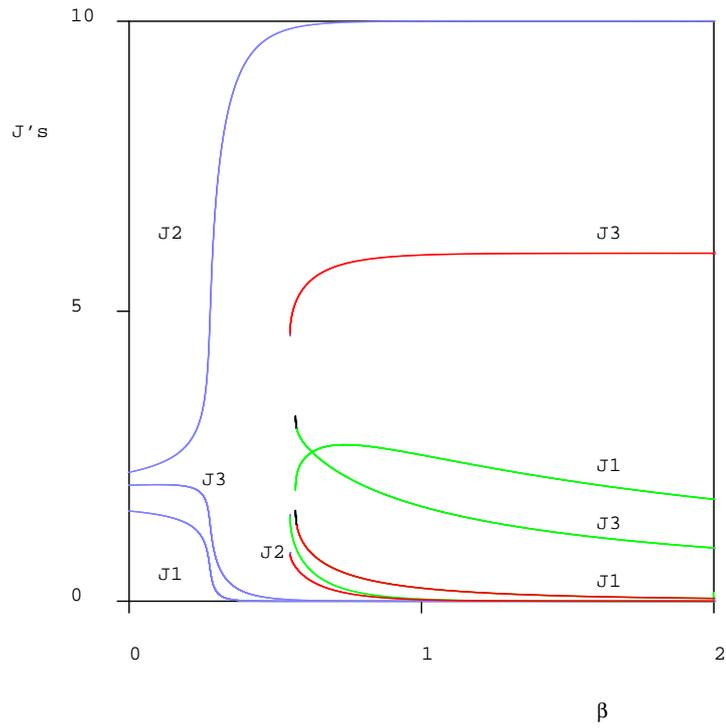} }
\caption{ Bifurcation diagram obtained by integration of the
mean field equations. $\gamma=0.05$ . (The continuation algorithm failed 
to converge to the exact position of the bifurcation around $\beta \approx 0.57$ ).
Attractor M is coloured blue, attractor L is coloured red
and attractor HL is coloured green. \cite{GRIND} software.
}
\end{figure}  

\FloatBarrier


\section{Agent-based simulations}

\subsection{Average analysis}

  Let us now compare the above results with those directly obtained by
  agent-based simulations.
   At each time step a pair of agents is randomly chosen.
   They play the bargaining game choosing their move with a probability given by equation (2) using their own specific $J_j$
    (not an average $J_j$ as in the mean field approximation),
   which they update after the session. And so on.
   
   We here report 4 types of results:
   \begin{itemize}
   \item On figure 3 the phase transition diagram (to be compared with figure 2). 
   \item On figure 4 individual trajectories in the $J$ simplex.
   \item On figure 5 the distribution of individual $J$'s on 4 attractors.
   \item On figure 6 a sketch of the attraction basins.    
   \end{itemize}   
  
We first monitor the different $J_j$ averaged over the whole population
at equilibrium when $\beta$ is scanned
downward and upward between 0 and 2 (figure 3). For the $\beta$ decreasing 
branches $J_{id}$, we start at $\beta=2$ from initial distributions of 
$J_j$ close to one of the attractors for each branch and carry on integration 
until the attractor is reached. The branches are continued when $\beta$
is lowered, taking as initial conditions the previous values of $J_j$ on the attractor.
The equivalent method is applied when $\beta$ is increased from 0
for the $J_{im}$ branches.
For the sake of clarity, attractors HL and attractor L are represented resp. on the upper and lower plots.

  Only attractor M can be reached
  when $\beta$ is increased from the disordered attractor.
  When  $\beta>0.571$, the path is reversible, but a hysteresis cycle is 
  observed when $\beta$ crosses the bifurcation.
  
   In the ordered region, the transition from attractor HL to attractor M
   is direct. By contrast, one first observes a continuous transition from L to HL around $\beta=0.75$ above the sharp transition at the bifurcation.
      
\begin{figure}
\epsfxsize=150mm\epsfbox{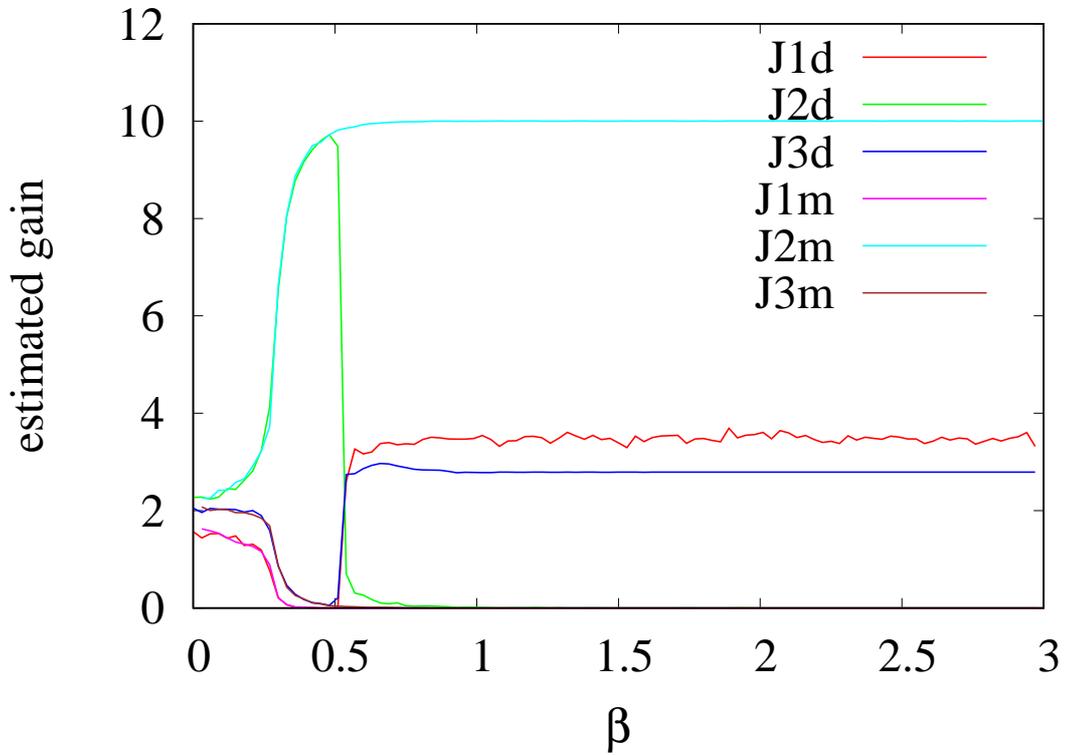}
\epsfxsize=150mm\epsfbox{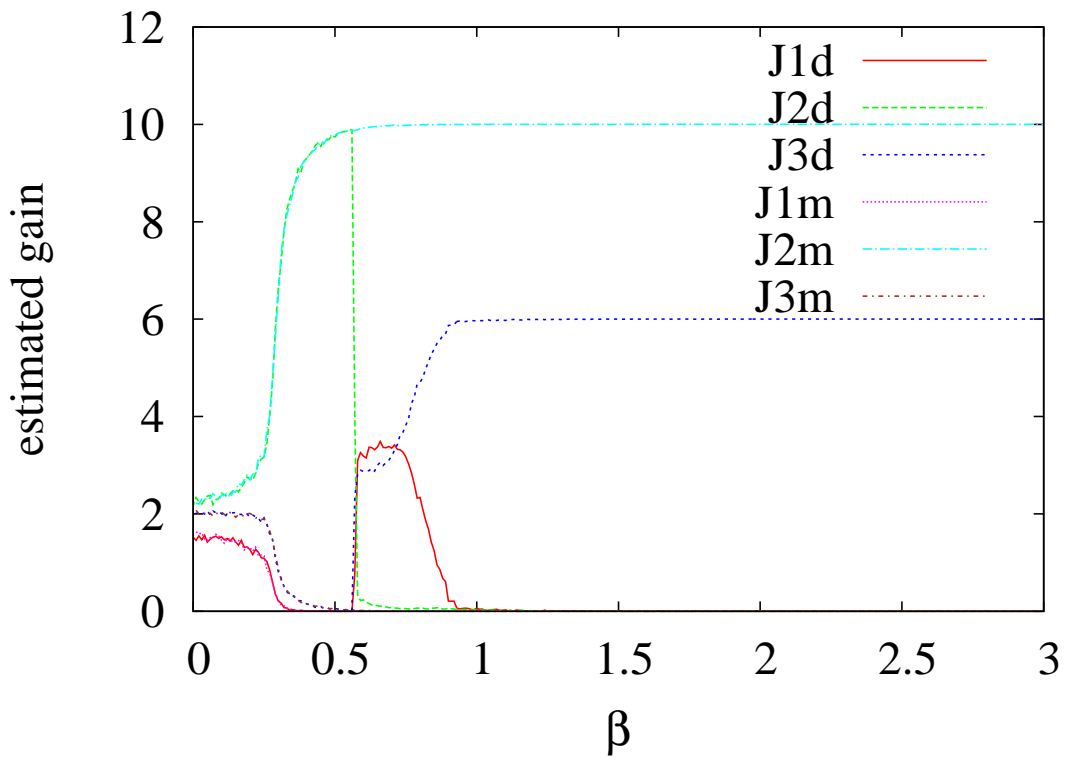} 
\caption{ Bifurcation diagrams obtained by AB simulations.
The six lines on each plot are values of the 3 preference coefficients J measured 
during $\beta$ decrease between 0 and 2 (J1d e.g.) and then increase (J1m e.g.) from
0 to 2. The decrease is started from the mixed attractor HL 
on the upper plot, and from the uniform L attractor on the lower plot.
Reversibility is only observed in the low $\beta$ or in the high $\beta$ region 
but not across the $\beta=0.571$ transition. 
}
\end{figure}


   Some attractor levels can be readily obtained from equilibrium
    conditions of equations (10-12).
  When only one move $j$ is chosen by the agents, the fraction
  involving exponentials equals 1 and the value of $J_j$ is given 
  by :
  \be
  J_j = \frac{\pi_i}{\gamma} 
  \nd
 in accordance with simulation results:
 on attractor M $J_2$ reaches $0.5/0.05=10$ and on
 attractor L $J_3$ reaches  $0.3/0.05=6$.
 In the case of the disordered attractor
 for low $\beta$ values, the exponentials are close to one 
 and the $J_j$ are directly computed from equations (10-12).
 
   Phase diagrams of the mean field approximation and of the 
   agent-based simulations look pretty similar with the same attractors;
   the main difference is the dependence of {\bf J} upon $\beta$
   for the HL attractor observed in the Mean Field Approximation.

 \subsection{Individual positions}
 
 The previous results concerned global features. Let us now examine
 individual agent choices. We use a simplex representation as in
 \cite{aey}. At any time step, preference coefficients J of an
 agent are displayed on the simplex by a point which position
 corresponds to the center of gravity of masses
 proportional to J1, J2, J3 placed at vertices H, M, L.
 For instance an agent positioned close to the center of the simplex is indifferent
 to choice H, M or L, while any agent close to one of the vertices has strong
 preferences and mostly plays according to that vertex.

  Figure (4) represents a typical set of 30 agents trajectories
  in the simplex for  $\beta=2$ and $\gamma=0.05$ during 10000 integration steps
  (each agent has been sampled 666 times on average).
 We started from a uniform distribution of initial $J$'s of width 2
 centered  around  (2.0, 2.0, 2.0). Their positions
 are indicated by a small square. 
 Trajectories are distinctively coloured.
 After a few initial wanderings, they diverge in the direction of the closest 
 vertex.  They remain fixed in the case of L and M vertices. 
 Some might fluctuate around vertex H because of possible encounters 
 with high demanding opponents which result in the decrease of $J_3$.  
      
\begin{figure} 
\centerline{\epsfxsize=120mm\epsfbox{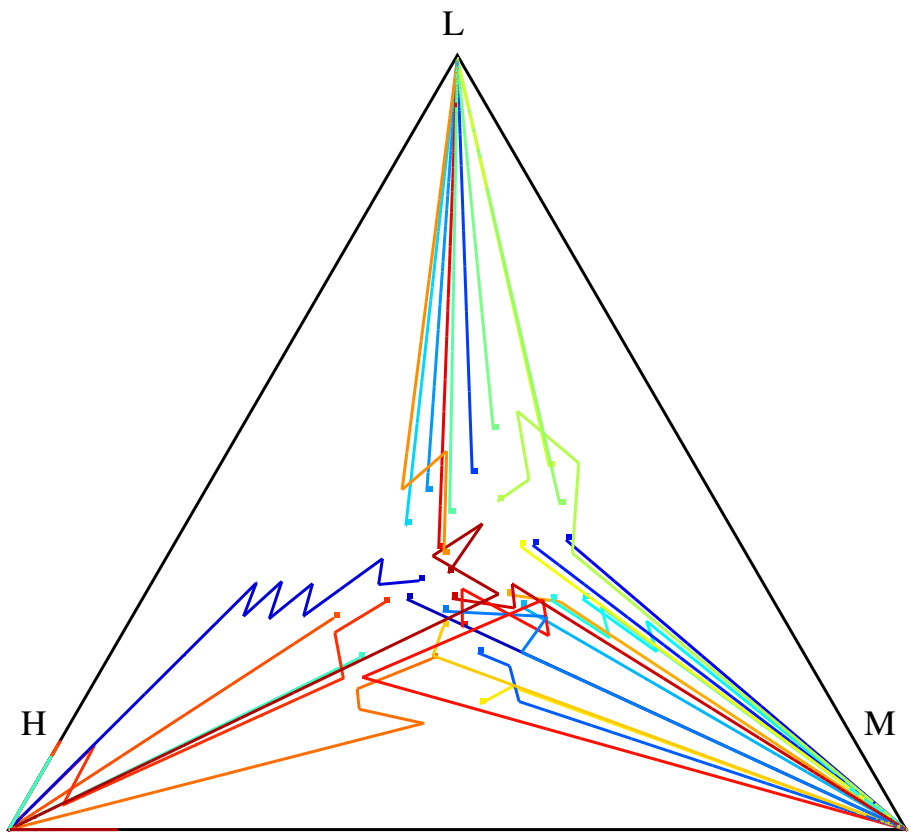} }
\caption{ Individual trajectories across the simplex.
Initial positions close to the center are indicated by little squares.
 $\beta=2$, $\gamma=0.95$, 10000 integration steps.}
\end{figure}

\vspace{1cm}

  Each of the 16 simplices on figure (5)
 is a snapshot of agents' 
 $J$s after a given integration time for a given value of $\beta$
 and for $\gamma=0.05$.
  
 Each red point represents the set of preference coefficients
 of a single agent.
 Each line of vertices displays the evolution of agents preferences
 at increasing iteration times towards one of the 4 asymptotic
 configurations for a given value of $\beta$.
 
   The initial conditions where chosen to favour the attractor
   to be displayed.  We used uniform distributions of width 1.0 around 
   (0.9, 0.9, 0.9) for the disordered attractor, D,  
  around (1.0, 0.6, 1.0) for the M attractor,
   around (1.0, 0.6, 3.0) for the HL attractor
   and around (0.6, 0.6, 6.5) for the L attractor.
 
 In agreement with previous observations, we see on the first line of 
 figure (5) that for low $\beta$
 values the agents positions remain dispersed inside the simplex
 even for long iteration times, which corresponds to a disordered phase.
 
  By contrast when $\beta$ is increased, agents 
  in the ordered phase gather towards one or two vertices,
  even after much smaller iteration times.
  We have chosen intermediate values of $\beta$ to avoid the accumulation of representative points on simplex vertices which would be observed 
  at larger $\beta$ values, e.g.  $\beta=2$.
        
    A physical interpretation of the above results
    would be a comparison with a condensed phase
   with thermal excitations above the ground state. When $\beta$ further
   increases, an equivalent of temperature decrease in physical systems,
   agents preferences condense exactly on the vertices (see further figure (6)),
   a property which helps us to check the basins of attraction. 
 
 \begin{figure}
\epsfxsize=120mm\epsfbox{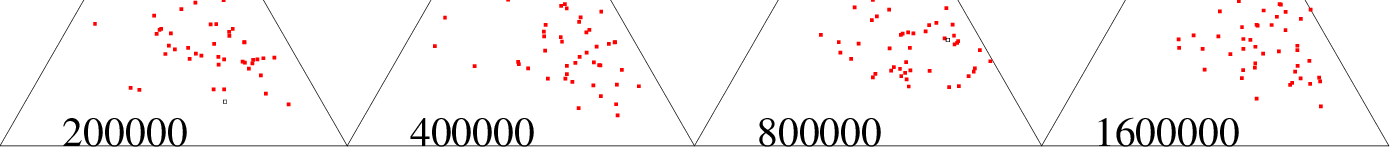}
\epsfxsize=120mm\epsfbox{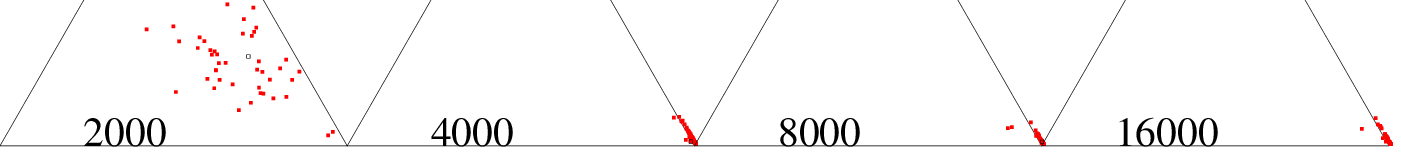} 
\epsfxsize=120mm\epsfbox{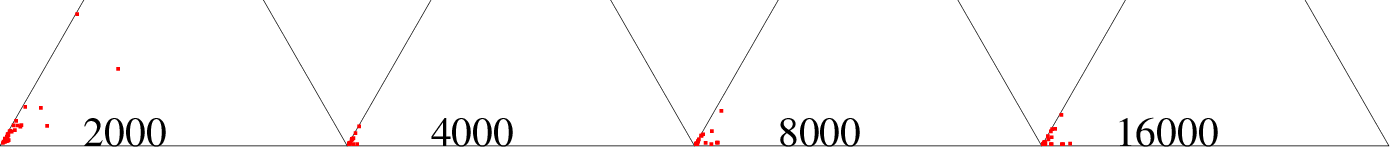}
\epsfxsize=120mm\epsfbox{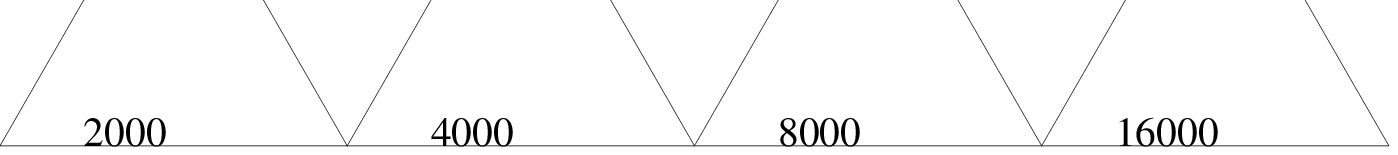} 
\caption{ Evolution of agents positions for different values of $\beta$.
$\gamma=0.05$. Each 4 simplex line represents the position of agents J's
after different iteration times written at the bottom of the simplex.
$\beta$ values are 0.2, 0.4 and twice 0.8 for the different lines.
Initial conditions are given in the text. 
}
\end{figure}

%

\subsection{Basins of attraction}

  The next question concerns the extension of the basins of attraction of the 
  different attractors. In fact, a systematic search for $\beta=2,\gamma=0.05$
  displays many more attractors than expected from our preliminary scans. 

  Figuring basins of attraction in a 3D phase space is not
  obvious and we once again use a vertex representation.
  The data are obtained by a triple scan of initial conditions.
  Each initial condition is a uniform distribution of J's values
  of width 1.0 around a given center. For instance, an initial distribution 
  centered on the center of gravity of the simplex is randomly 
  drawn in the cube:
  $2.7 < J_1 < 3.7\, ,  \,   2.7 < J_2 <3.7 \, ,  \, 2.7 < J_3 < 3.7 $.
   The upper circle of figure 6 corresponds to the initial distribution: $0.3 < J_1 < 1.3 \, ,  \, 0.3  < J_2 < 1.3 \, ,  \, 2.7 < J_3 < 3.7 $ etc.
  
  Figure (6) describes data gathered by 3 nested loops across
  initial $J_1, J_2, J_3$,
  where distribution centers are varied from 0.4 to 6.4 by a factor 2.
  The positions of the centers of the circles in the simplex codes the initial distribution centers. We used the condensation of agents preferences 
  on the 3 vertices to display final distributions by pie charts. 
   Red sectors represent the percentage of agents with choice H,
blue sectors represent the percentage of agents with choice M,
green sectors represent the percentage of agents with choice L.
Rare and narrow white sectors represent the percentage of agents 
inside the simplex. We checked that they are located
 in the immediate neighbourhood of vertices.

\begin{figure}
\hspace{-1.5cm}
\rotatebox{-90}{
\epsfxsize=120mm\epsfbox{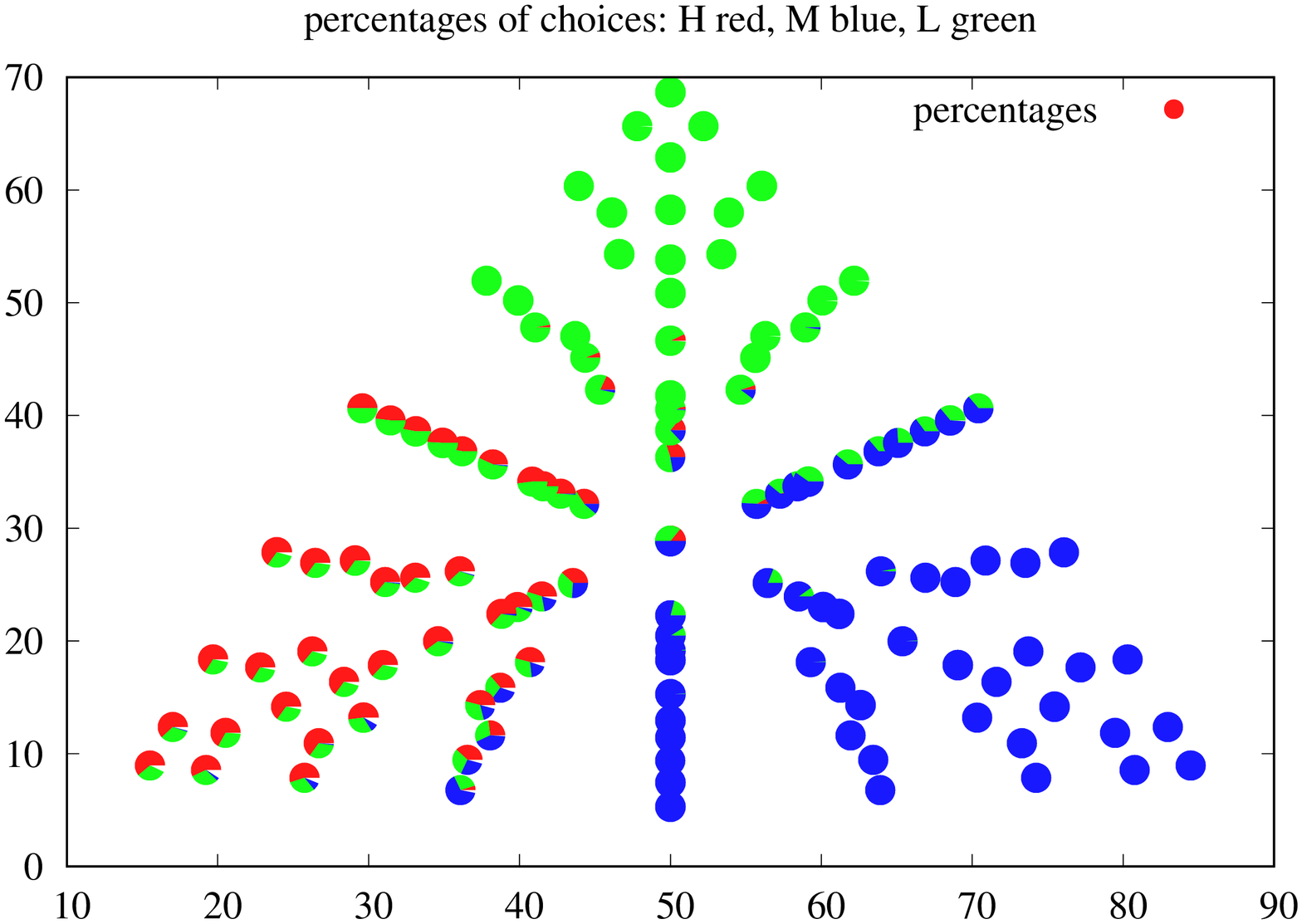}}
\caption{Simplex representation of the basin of attractions.
The position of the circles in the simplex codes the initial conditions.
Each circle codes the percentage of agents who reached the 3 vertices.
Red sectors represent the percentage of agents with choice H,
blue sectors represent the percentage of agents with choice M,
green sectors represent the percentage of agents with choice L.
Rare and narrow white sectors represent the percentage of agents 
inside the simplex.
$\beta=2.0$, $\gamma=0.05$, 100 agents, 100000 iteration steps.
}
\end{figure}

  Several important conclusions about ordered phases for large $\beta$ can be drawn:
  \begin{itemize}
  \item Mixed strategies inside the vertex are unstable.
  \item The attractors are distributions of agents on the vertices
  or very close. Agents have strong opinions about the value of their choice and don't change them frequently. 
  \item J space is paved with basins of attraction surrounding the attractors. The dynamics collapse choices to the nearest vertex,
  with the exception of vertex H.
  \item No distribution consists of only H preferences - which would 
   give no gain to the agents. When the initial conditions are close to vertex H,
  the attractors can only be mixed distributions, with very few agents playing M. 
  \item \citep{aey} and \citep{poza} report the existence of "fractious attractor"
  such that agents oscillate between choices H and L for long transient.
  We never observed such "fractious attractor", even after a specific search,
  and we suspect that they were
  due to their choice of a constant noise term. 
  \end{itemize}
  


 
\section{Tagged populations}

  The most striking result in \cite{aey} is the existence 
  of an inequity norm sustainable among two {\it a priori} equivalent
  tagged groups. Their inequity norm corresponds in our settings
  to an attractor such that all members of one population with tag T1
  play H against any member of the other tagged population (T2) who 
  always plays L against them. 
  
  To investigate the basins of attractions for the two tagged populations 
  we proceed with the same scan of initial conditions of population with tag T2
  as above, but maintain the same initial conditions for population with tag T1
  around the center of the simplex:
 
  $2.5 < J_1 < 3.5\, ,  \,   2.5 < J_2 <3.5 \, ,  \, 2.5 < J_3 < 3.2 $.  
   
  Figure 7 displays the attractors of the inter-population dynamics, the simplex T1 above simplex T2 with the same conventions as in figure 5, except that the positions on both simplices correspond to the scan of initial conditions of population T2.
The colour codes are the same as for figure 6 and reflects the attractors of each population. 

The pie charts close to the left vertices of the simplices e.g. are coloured
green for T1 and red for T2; the attractors of the dynamics are then 
L for T1 and H for T2.
 The pie charts close to the top vertices correspond 
to a stable mixture of the 3 possible moves H,M,L for T1 and to pure L for T2 
The pie charts close to the right vertices correspond 
to a stable mixture of 2 possible moves M,L for T1 and to pure M for T2.

  Not surprisingly, the attractors reflect the initial conditions
  of population T2 since the initial conditions of T1 were kind of neutral. 
  
   The main difference with the no-tag simulation is the 
   appearance of a pure strategy attractor such that all agents with tag T2 play H against agents with tag T1 who play L. This asymmetrical attractor parallel the findings of \citep{aey} who refer to a "discriminatory norm". But our analysis 
   characterises an attractor reached from initial conditions that were 
   already biased towards inequity with larger values of $J_1$. And this 
   condition is an attractor of the dynamics, not a transient.


\begin{figure}
\hspace{-9cm}
\rotatebox{-90}{
\epsfxsize=210mm\epsfbox{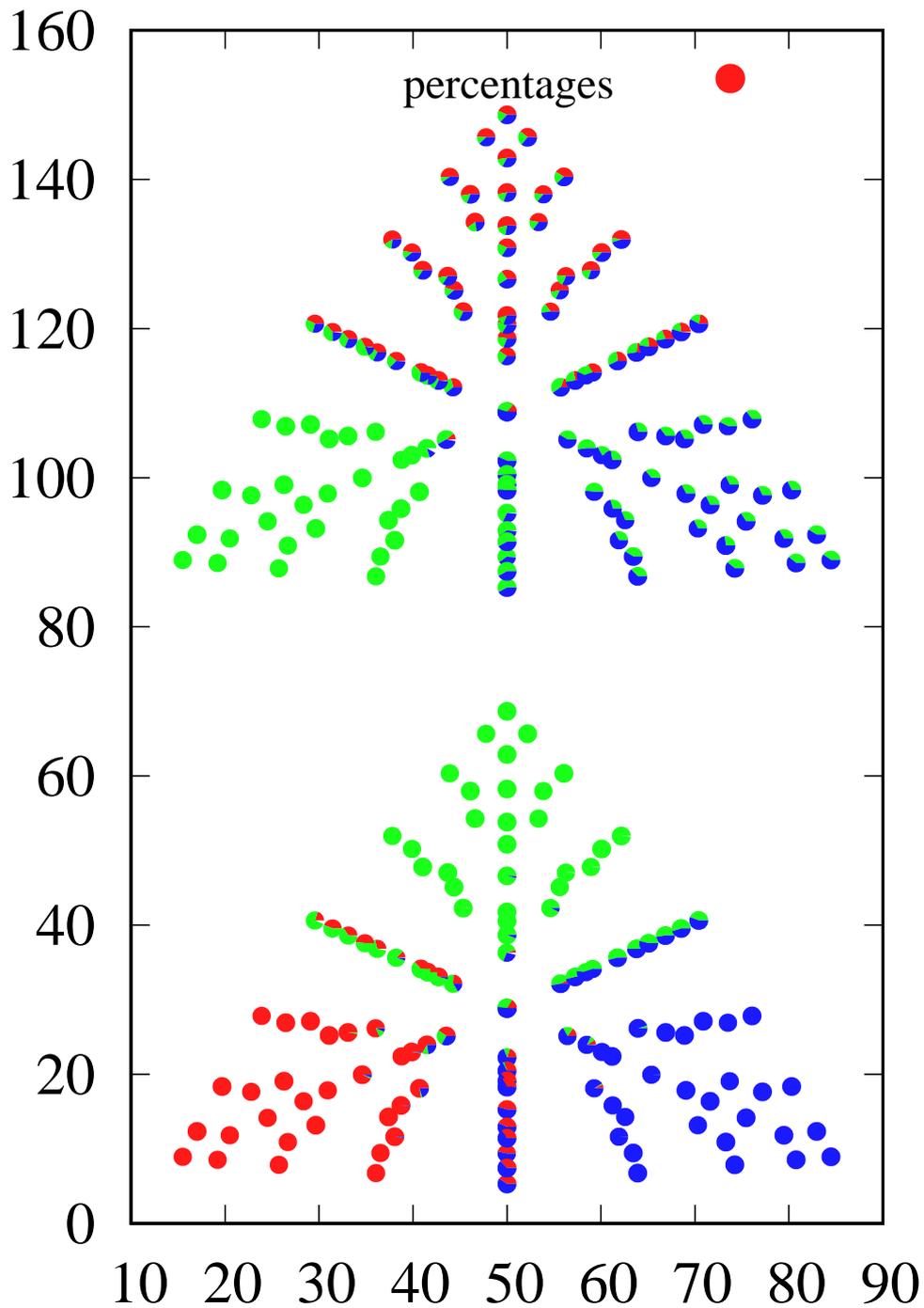}}
\caption{Simplex representations of the basin of attractions for
two tagged populations of 100 agents,
 one with tag T1 above and one with tag T2 below.
As previously, colours code the attractors of the dynamics
with the same coding as in figure 6.
Positions in the simplex code the initial conditions of players 
with tag T2 \emph{for both groups} while
initial conditions of tag T1 players remain identical around the center
of the simplex.
$\beta=2.0$, $\gamma=0.05$, 100 000 iteration steps
per data.
}
\end{figure}

\section{Discussion and Conclusions}

  The reformulation of the iterated bargaining game of \citep{aey}
  using more elaborate cognitive processes such as taking
moving averages of past gains and choosing next moves according to
Boltzman probabilities allows a more precise description of the dynamics
in terms of attractors, regime transitions and basins of attraction. The number of parameters was reduced from two to one. Several transitions are observed between 
a disordered state and several stable ordered configurations when $\beta$
increases. Because
we use Boltzman choice function, agents end-up using mostly pure strategies
for larger values of $\beta$. We never observed any "fractious state"
such that agents remain in the interior of the simplex changing
randomly their choice between H and L as reported in \citep{aey} and \citep{poza}.
 Our guess is that such behaviour is due to their hypothesis
 of a random choice with a constant non-zero probability.
 
 As discussed earlier in sections 4.3 and 5, {\textbf{J}} space is paved with basins of attraction surrounding the attractors and dynamics collapse preference coefficients to the nearest vertex.  Unbiased random initial conditions never generate H/L attractors.

 Hence our interpretation in terms of social phenomena: game interactions and cognitive processes can increase and stabilise discrimination and inequality among
   tagged populations, even when tag were {\it a priori} neutral. On the other hand, inequality attractors never "emerge"\footnote{\citep{aey} specify in a footnote
 that they "use the term "emergent" ... to mean simply "arising
from the local interactions of agents." ". But the word Emergence in the title
of their paper evokes the idea of emergence
 of Classes in a previously egalitarian society.
} spontaneously from random unbiased initial conditions. 

The changes we introduced in \citep{aey} also allow to figure out which properties 
can be considered as generic, that is to say independent from the details of each model,
and which are specific to the exact formulation of the model.
   
   The two versions of the iterated bargaining game,
   \citep{aey} and the present paper, agree that unfair
   social institutions such as classes and discrimination can result
   as the downside of a rational cognitive practice, namely memorising
   or coding previous events to take present decisions. And furthermore, that taking into account {\it a priori} irrelevant tags can lead
   to a dissociation of the two tagged populations into an upper and a lower class.  
   
   But we differ by our interpretation in terms of social phenomena: game interactions and cognitive processes can increase and stabilise discrimination, but inequality attractors never "emerge" spontaneously from random unbiased initial conditions. History has taught us that wars and invasions often result into discriminations
   that are maintained long after these events.

   \vspace{1cm}

{\bf Acknowledgments}

We thank Sophie Bienenstock, Bernard Derrida, Alan Kirman, Jean-Pierre Nadal
and Jean Roux for helpful discussions and David Poza for providing his Netlogo
program of the lattice version of \citep{aey} model. 


\bibliography{biblio}

\end{document}